\documentclass[12pt]{iopart}

\usepackage{iopams}
\begin{document}

\title{Holographic dark energy and late cosmic acceleration}

\author{Diego Pav\'{o}n}

\address{Departmento de F\'{\i}sica, Universidad Aut\'{o}noma de  Barcelona, 08193 Bellaterra, Spain}
\ead{diego.pavon@uab.es}

\begin{abstract}
It has been persuasively argued  that the number of the effective
degrees of freedom of a macroscopic system is proportional to its
area rather than to its volume. This  entails interesting
consequences for cosmology. Here we present a model based on this
``holographic principle" that accounts for the present stage of
accelerated expansion of the Universe and significantly alleviates
the coincidence problem also for non-spatially flat cosmologies.
Likewise, we comment on a recently proposed late transition to a
fresh decelerated phase.
\end{abstract}

\maketitle

\section{Introduction}
Nowadays there is an ample consensus, deeply rooted in
observational grounds, that the Universe is currently undergoing a
phase of accelerated expansion likely driven for some field
(dubbed ``dark energy field") that clusters, if any, only at the
largest accessible scales, able to generate a negative pressure
large enough to violate the strong energy condition -see
\cite{consensus} and references therein. By far, the conceptually
simplest dark energy candidate is the cosmological constant,
$\Lambda$. Albeit thus far it fits reasonably well all the
cosmological data it confronts two serious drawbacks on the
theoretical side. On the one hand, its quantum field value results
about 123 orders of magnitude larger than observed. On the other
hand, it gives rise to the {\em coincidence problem}, namely:
``why are the vacuum and dust energy densities of precisely the
same order today?" (Bear in mind that the energy density of dust
red-shifts with expansion as $a^{-3}$, where $a$ denotes the scale
factor of the Robertson--Walker metric). This is why a number of
candidates  of varying degree of plausibility have been proposed
over the last years with no clear winner in sight -see
\cite{edmund} for a recent review. Here we focus on a dark energy
candidate grounded on sound thermodynamic considerations that is
receiving growing attention in the literature, namely, the
``holographic dark energy".

\section{Holographic dark energy}
We begin by briefly introducing the holography concept  after 't
Hooft \cite{hooft} and Susskind \cite{susskind}. Consider the
world as three-dimensional lattice of spin-like degrees of freedom
and assume that the distance between every two neighboring sites
is some small length $\ell$. Each spin can be in one of two sates.
In a region of volume $L^{3}$ the number of quantum states will be
$N(L^{3}) = 2^{n}$, with $n= (L/\ell)^{3}$ the number of sites in
the volume, whence the entropy will be $S \propto (L/\ell)^{3} \ln
2$. One would expect that if the energy density does not diverge,
the maximum entropy varies as $L^{3}$,
i.e., $S \sim L^{3} \, \Lambda^{3}$, where $\Lambda \equiv %
\ell^{-1}$ is to be identified with the ultraviolet cutoff.
However, the energy of most states so described  would be so big
that they will collapse to a black hole larger than $L^{3}$. It
seems therefore reasonable  that  in the quantum theory of gravity
the maximum entropy should be proportional to the area, not the
volume, of the system under consideration. (Recall that the
Bekenstein--Hawking entropy is $S_{BH} = A/(4 \,%
\ell_{Pl}^{2})$, where $A$ is the area of the black hole horizon).

Consider now a system of volume $L^{3}$ of energy slightly  below
that of a black hole of the same size but with entropy larger than
that of the black hole. By tossing into the system a tiny amount
of energy a black hole would result but with smaller entropy than
the original system thus violating the second law of
thermodynamics. As a consequence, Bekenstein suggested that the
maximum entropy of the system should be proportional to its area
rather than to its volume \cite{jakob}. In the same vein `t Hooft
conjectured that it should be possible to describe all phenomena
within a volume by the set of degrees of freedom residing on its
boundary. The number of degrees of freedom should not exceed that
of a two-dimensional lattice with about one binary degree of
freedom per Planck area.

Inspired in these ideas, Cohen {\it et al.} \cite{cohen} argued
that an effective field theory that saturates the inequality
$L^{3}\, \Lambda^{3} \leq S_{BH}$ necessarily includes many states
with $R_{s} > L$, where $R_{s}$ is the Schwarzschild radius of the
system under consideration. It seems therefore  reasonable to
propose a stronger constraint on the infrared cutoff $L$ that
excludes all states lying within $R_{s}$, namely, $L^{3}\,%
\Lambda^{4} \leq m_{Pl}^{2}\, L$ (clearly, $\Lambda^{4}$ is the
zero--point energy density associated to the short-distance
cutoff). So, we may conclude that $L\sim \Lambda^{-2}$ and
$S_{max} \simeq S_{BH}^{3/4}$. By saturating the inequality -which
is not compelling at all- and identifying $\Lambda^{4}$ with the
holographic dark energy density one has \cite{Li}
\\
\begin{equation}
\rho_{x} = 3 c^{2}\, M_{p}^{2}/L^{2} \qquad \quad (M_{p}^{2}%
\equiv (8\pi G)^{-1}),
\label{rhox}
\end{equation}
where $c^{2}$ is a dimensionless constant.

Suggestive as they are, the above ideas provide no indication
about how to choose the infrared cutoff in a cosmological context.
Different possibilities have been tried with varying degrees of
success, namely, the particle horizon \cite{particleh}, the future
event horizon \cite{Li, fevh} and the Hubble horizon \cite{plb628,
non-flat}. Here we shall adhere to the latter for it looks the
most natural one.

\section{Interacting dark energy}
Our model rests on three main assumptions: $(i)$ the dark energy
density is given by Eq. (\ref{rhox}), $(ii)$ $L = H^{-1}$, where
$H \equiv \dot{a}/a$ is the Hubble function,  and ($iii$) matter
and holographic dark energy do not conserve separately but the
latter decays into the former with rate $\Gamma >0 $, i.e.,
\begin{equation}
\dot{\rho}_{m} + 3H\rho_{m} = \Gamma\, \rho_{x}\, , \; \qquad
\qquad \dot{\rho}_{x} + 3H(1+w)\rho_{x} = - \Gamma \rho_{x}\, .
\label{balance}
\end{equation}
In spatially flat universes there is a relation connecting the
equation of state parameter of the dark energy to the ratio
between the energy densities, $r \equiv \rho_{m}/\rho_{x}$, and
$\Gamma$, namely, $w = -(1+r)\Gamma/(3rH)$, such that any decay of
the dark energy into pressureless matter implies a negative $w$.
It also follows that the ratio of the energy densities is a
constant, $r_{0} = (1-c^{2})/c^{2}$, whatever $\Gamma$ -see Ref.
\cite{plb628} for details.

In the particular case that $\Gamma \propto H$ one has $\rho_{m}$,
$\rho_{x}$ $\propto a ^{-3m}$ and $a \propto t^{n}$ with $m =
(1+r_{0}+w)/(1+r_{0})$ and $n = 2/(3m)$. Hence, there will be
acceleration for $w < -(1+r_{0})/3$. In consequence, the
interaction is key to simultaneously solve the coincidence problem
and have late acceleration. For $\Gamma = 0$ the choice $L =%
H^{-1}$ does not lead to acceleration. We wish to emphasize that
models in which matter and dark energy interact with each other
considerably alleviate the coincidence problem \cite{interacting}
and fare remarkably well when measured against observational data
\cite{gmo}.

Obviously, prior to the current epoch of accelerated expansion a
matter dominated period is required for the standard picture of
cosmic structure formation to hold. The usual way to incorporate
this is to assume that the ratio $r$ has not been constant but was
(and possibly still is) decreasing towards some final value
$r_{0}$. In the present context, a time dependence of $r$ can only
be achieved by allowing the parameter $c^{2}$ to slowly vary with
time. By ``slowly" we mean that $\left(c^{2}\right)^{\displaystyle
\cdot}/c^{2} \ll H $. This is not only permissible but reasonable
since it is natural to expect that the holographic bounds gets
fully saturated only in the very long run or even asymptotically.
Our approach, however, offers a different way to recover an early
matter dominated epoch. Namely, for $\Gamma/H  \ll 1$, the dark
energy itself behaves as pressureless matter since one has $\mid w
\mid \ll 1$, even for a constant $r$. From the definition of
$\rho_{x}$ it follows that
\\
\begin{equation}
\dot{\rho}_{x}  = - 3 H \, \left[1 + \frac{w}{1 + r}\right]\,
\rho_{x}+ \frac{\left(c^{2}\right)^{\displaystyle \cdot}}{c^{2}}\,
\rho_{x}\ . \label{dotrhox1}
\end{equation}
Combining this with the balance equation (\ref{balance}.b) and
contrasting the resulting expression with the evolution equation
for $r$,
\\
\begin{equation}
\dot{r} = 3 H r \left[w + \frac{1 +
r}{r}\frac{\Gamma}{3H}\right]\, ,
\label{dotr}
\end{equation}
yields $\left(c^{2}\right)^{\displaystyle \cdot}/c^{2} = -%
\dot{r}/(1+r)$, whose solution reads $ c^{2}(t) =
\frac{1}{1+r(t)}$ . At late times, $r \rightarrow r_{0}$ whence
$c^{2} \rightarrow c_{0}^{2}$. In this scenario $w$ depends also
on the fractional change of $c^{2}$ according to
\\
\begin{equation}
w = - \left(1 +  \frac{1}{r}\right) \left[\frac{\Gamma}{3 H} +
\frac{\left(c^{2}\right)^{\displaystyle \cdot}}{3 H c^{2}}
\right]\, .
\label{wx}
\end{equation}
Since the holographic dark energy must satisfy the dominant energy
condition (and therefore it is not compatible with ``phantom
energy" \cite{bak}), the restriction $w \geq -1$ sets constraints
on $\Gamma$ and $c^{2}$.

For future convenience we write the deceleration parameter
\begin{equation}
q = \frac{1}{2}\Omega_{m} + \frac{1}{2}(1+3w)\Omega_{x} \, ,
\label{decelp}
\end{equation}
where $\Omega_{m}$ and $\Omega_{x}$ stand for the dimensionless
density parameters of matter and dark energy, respectively. Up to
now we have restricted our attention to spatially flat
Friedmann-Lemaitre-Robertson-Walker (FLRW) universes. It proves
illustrating to extend the study to FLRW models with curved
spatial sections.

\subsection{FLRW universes with $k \neq 0$}
Aside from the sake of generality other motivations for models
with non-flat spatial sections are: $(i)$ Inflation drives the
$k/a^{2}$ ratio close to zero but it cannot set it to zero if $k
\neq 0$ initially. $(ii)$ The closeness to perfect flatness
depends on the number of $e$-folds and  we can only speculate
about the latter. $(iii)$ After inflation the absolute value of
the $k/a^{2}$ term in Friedmann's equation is bound to steadily
increase with respect to the matter density term, thereby the
former should not be ignored when studying the late Universe.

For curved spatial sections, Eqs.  (\ref{dotr}) and (\ref{wx})
generalize to
\begin{equation}
\dot{r} = - 3H\,r\,\frac{1}{1 -
\Omega_{x}}\,\left\{\frac{k}{a^{2}H^{2}}\,\left[\frac{1}{r}\,
\frac{\Gamma}{3H}- \frac{1}{3}\right] +
\frac{1}{3H}\,\frac{\left(c^{2}\right)^{\displaystyle
\cdot}}{c^{2}} \right\}\ ,\label{rdot1}
\end{equation}
and
\begin{equation}
w  = - \frac{1}{1 - \Omega_{x}}\,\left[\frac{\Gamma}{3H} -
\frac{1}{3}\,\frac{k}{a^{2}H^{2}} + \frac{1}{3H}\,
\frac{\left(c^{2}\right)^{\displaystyle \cdot}}{c^{2}}\right]\
,\label{wx1}
\end{equation}
respectively -see \cite{non-flat} for details. We see that, aside
from the evolution of $c^{2}$, the evolution of the matter-dark
energy ratio $r$ is immediately connected to a non-vanishing
spatial curvature which may help to speed the decrease of the
former. On the other hand, because
$\left(c^{2}\right)^{\displaystyle \cdot}/c^{2} \ll H $ by
assumption and $\mid k/(aH)^{2}\mid_{0} \ll 1$ by observation it
follows  that $\mid\dot{r}/r\mid_{0} \ll H_{0}$, i.e., the
coincidence problem gets greatly alleviated (bear in mind that in
the conventional $\Lambda$CDM scenario $\mid\dot{r}/r\mid_{0} =
3H_{0}$).

Likewise, the curvature term modifies the equation of state
parameter $w$. Depending on the whether the Universe is spatially
open or closed the negative character of $w$ will be accentuated
or softened. A detailed analysis of the impact of the curvature
term on this and related issues as the transition from
deceleration to acceleration can be found in Ref. \cite{non-flat}.

\section{Transition to a new decelerated era?}
It has been speculated that the present phase of accelerated
expansion is just transitory and that the Universe will eventually
revert to a fresh decelerated era. This can be achieved by taking
as dark energy a scalar field whose energy density obeys a
suitable ansatz. As a result the equation of state parameter $w$
evolves from values above but close to $-1$ to much less negative
values thereby the deceleration parameter increases to positive
values \cite{carvalho}. Thus, the troublesome event horizon that
afflicts superstring theories disappears altogether. Here we shall
argue that our holographic interacting model -which was devised to
provide a transition from deceleration to acceleration and
alleviate the coincidence problem- is in principle  compatible
with such a transition.

For the sake of simplicity we set $k = 0$. Inspection of Eq.
(\ref{wx}) reveals that $w$ can become larger than $-1/3$ (which
by Eq. (\ref{decelp}) means deceleration) either by allowing any
of the two terms in the square parenthesis, or both, to reach
sufficiently small values or just keeping the first term nearly
constant and allowing the second one to become negative enough.
Clearly, all these possibilities look a bit contrived, especially
the latter one as -contrary to intuition-, in such a case,  the
saturation parameter does not increase but decreases. However, we
should no wonder at this as the proposal of coming back to a
decelerated phase for the sole purpose of getting rid of the event
horizon appears rather artificial, especially because nothing in
the observational data hints at that. Nonetheless, we should keep
an open mind since this possibility cannot be dismissed offhand.
At any rate, we wish to emphasize that those holographic dark
energy models that identify the infrared cutoff $L$ with the event
horizon are unable to account for such a transition.

In all, the holographic dark energy provides a simple and elegant
thermodynamic-based explanation, within Einstein relativity, for
the present era of cosmic accelerated expansion. Moreover, it
substantially alleviates the coincidence problem provided that
matter and dark energy do not conserve separately. The model can,
in principle, accommodate a later transition to a new decelerated
phase.

\section*{Acknowledgments}
Thanks are due to the organizers of the second edition of the
IRGAC series of Conferences. I am greatly indebted to Winfried
Zimdahl for conversations and suggestions on the subject of this
work. This research was partially supported by the Spanish
``Ministerio de Educaci\'{o}n y Ciencia" under Grant
FIS2006-12296-C02-01.


\section*{References}
\begin{thebibliography}{99}
\bibitem{consensus} Padmanabhan T  2003 Phys Reports \textbf{380} 235-320\\
Sahni V 2004 astro-ph/0403324\\
Lima JAS 2004 Braz. J. Phys. \textbf{34} 194-200\\
Riess AG {\it et al} 2004 Astrophys. J. \textbf{607} 665-687\\
L. Perivolaropoulos 2006 astro-ph/0601014\\
Astier P, {\it et al} 2006 Astron. \& Astrophys. \textbf{447} 31-48\\
Spergel DN, {\it et al} 2006 astro-ph/0603449
\bibitem{edmund}
Copeland E, Sami M and Tsujikawa S 2006 hep-th/0603057
\bibitem{hooft} `t Hooft G 1993 gr-qc/9311026
\bibitem{susskind} Susskind L 1995  J. Math. Phys. (NY) \textbf{36}
6377-6396
\bibitem{jakob} Bekenstein JD 1994 Phys. Rev. D \textbf{49}
1912-1921
\bibitem{cohen} Cohen AG, Kaplan DB and Nelson AE 1999 Phys. Rev.
Lett. \textbf{82} 4971-4974
\bibitem{Li} Li M 2004 Phys Lett B \textbf{603} 1-5
\bibitem{particleh} Fischler W and Susskind L 1998
hep-th/9806039\\
Cataldo M, Cruz N, del Campo S and Lepe S 2001 Phys. Lett. B
\textbf{509} 138-142
\bibitem{fevh} Guberina B, Horvat R and Nikolic H 2005
Phys. Rev. D \textbf{72} 125011(6)\\
Wang B, Gong Y and Abdalla E 2005 Phys. Lett B
\textbf{624} 141-146\\
Huang Q-G and Li M 2004 JCAP08(2004)013\\
Gong Y, Wang B and Zhang Y-Z 2005 Phys. Rev. D \textbf{72} 043510(6)\\
Nojiri S and Odintsov SD 2006 Gen. Rel. Grav. \textbf{38}
1285-1304\\
Wang B, Lin C-Y and Abdalla E 2006 Phys. Lett. B \textbf{637}
357-361
\bibitem{plb628} Pav\'{o}n D and Zimdahl W 2005 Phys. Lett. B
\textbf{628} 206-210
\bibitem{non-flat} Zimdahl W and Pav\'{o}n D 2006 astro-ph/0606555
\bibitem{interacting}Amendola L 2000 Phys. Rev. D \textbf{62}
043511(10)\\
Chimento LP, Jakubi AS, Pav\'{o}n D and Zimdahl W 2003 Phys. Rev.
D \textbf{67} 083513(11)\\
del Campo S, Herrera R, Olivares G and Pav\'{o}n D 2006 Phys. Rev.
D \textbf{74} 023501(9)
\bibitem{gmo} Olivares G, Atrio-Barandela F and Pav\'{o}n D 2005
Phys. Rev. D \textbf{71} 063523(7)\\
Olivares G, Atrio-Barandela F and Pav\'{o}n D 2006 \textbf{74}
023501(9)
\bibitem{bak}
Bak D and Rey S-J 2000 Class. Quantum Grav. \textbf{17} L83-l89
\bibitem{carvalho} Carvalho FC, Alcaniz JS, Lima JAS and Silva R
2006 Phys. Rev. Lett. \textbf{97} 081301(4)\\
Alcaniz JS 2006 astro-ph/0608631
\end{thebibliography}
\end{document}